# Tuning of Silicon Nitride Micro-Cavities by Controlled Nanolayer Deposition

Dmitry A. Kalashnikov[1,*], Gandhi Alagappan[2], Ting Hu[3], Nelson Lim[1], Victor Leong[1], Ching Eng Png[2] and Leonid A. Krivitsky[1,*]

[1]*Institute of Materials Research and Engineering, Agency for Science, Technology, and Research (A*STAR), 2 Fusionopolis Way, #08-03 Innovis, 138634 Singapore*
[2]*Institute of High Performance Computing, Agency for Science, Technology, and Research (A*STAR), Fusionopolis, 1 Fusionopolis Way, #16-16 Connexis, Singapore 138632*
[3]*Institute of Microelectronics, Agency for Science, Technology, and Research (A*STAR), 2 Fusionopolis Way, #08-02 Innovis, Singapore 138634*
\* Dmitry_Kalashnikov@imre.a-star.edu.sg and Leonid_Krivitsky@imre.a-star.edu.sg

**Abstract:** Integration of single-photon emitters (SPEs) with resonant photonic structures is a promising approach for realizing compact and efficient single-photon sources for quantum communications, computing, and sensing. Efficient interaction between the SPE and the photonic cavity requires that the cavity's resonance matches the SPE's emission line. Here we demonstrate a new method for tuning silicon nitride ($Si_3N_4$) microring cavities via controlled deposition of the cladding layers. Guided by numerical simulations, we deposit silicon dioxide ($SiO_2$) nanolayers onto $Si_3N_4$ ridge structures in steps of 50 nm. We show tuning of the cavity resonance over a free spectral range (FSR) without degradation of the quality-factor (Q-factor) of the cavity. We then complement this method with localized laser heating for fine-tuning of the cavity. Finally, we verify that the cladding deposition does not alter the position of nanoparticles placed on the cavity, which suggests that our method can be useful for integrating SPEs with photonic structures.

## 1. Introduction

Integrated quantum photonic devices are considered critical elements for future quantum networks, elements of quantum computers, and sensors [1-8]. One of the essential elements of such a device is a single photon emitter (SPE), which is coupled to a high Q-factor cavity. Upon excitation of the SPE, the single photon is emitted in the cavity mode and then routed to an optical network for manipulation and detection [9-12]. Such an interface requires a nearly perfect matching of the emission line of the SPE with the resonance line of the cavity. Given the intrinsic uncertainty in the SPE emission line and the cavity's resonances, their matching requires active tuning of the SPE and/or the cavity. While tuning the emission wavelengths of the SPE is possible, for example, by applying electric fields [13-15], it is arguably more practical to tune the resonance line of the photonic cavity.

Methods of tuning micro-cavities include thermal and electro-optical tuning, application of mechanical stress, and functionalization of the surface [16-26]. These methods allow accurate tuning of the cavity resonances with real-time control. However, they are not free from technical challenges. For example, thermal and electrical tuning require a significant amount of electrical power to be delivered into the chip, especially when broadband tuning across a full free spectral range (FSR) is required [18-20, 23-26]. Moreover, in this case, the electrical contacts should be positioned next to the cavities, which complicates the fabrication process and might degrade the cavity performance [18]. Methods based on the deposition of a photochromic film or the functionalization with a laser addressable polyelectrolyte strongly depend on the thickness of the applied material. These methodsmay also degrade the optical properties of the cavities due to surface modification [16, 17]. Stress-based approaches require sophisticated fabrication and a large device size to tune across the full FSR [21, 22].

Silicon nitride (Si$_3$N$_4$, or SiN) stands out as a material of choice for many device prototypes due to its CMOS compatibility, broad transparency range, and relatively high refractive index (n~2) [27-30]. However, the thermo-optic coefficient of Si$_3$N$_4$ is approximately one order of magnitude lower than for silicon [31], which makes it challenging to implement thermal-based tuning strategies described above. Moreover, at cryogenic temperatures, which is the optimal operation range for many SPEs, the Si$_3$N$_4$ thermo-optic coefficient becomes even smaller.

In this work, we propose and realize a method for broadband tuning of Si$_3$N$_4$ cavities. It is based on depositing a silicon dioxide (SiO$_2$) cladding over Si$_3$N$_4$ cavities with tens of nanometers step size. We show that within some range of thicknesses of the cladding layer, we can tune the resonance position within one FSR. Although this method can only tune the resonance in relatively coarse steps, it can be complemented with other methods for fine tuning. Nevertheless, this method allows us to first bring the cavity resonances and SPE emission wavelength closer to a rough approximation, before fine-tuning with an another approach. To this end, we demonstrate fine tuning of the cavity by applying laser heat after depositing the SiO$_2$ layer. We also show that SiO$_2$ deposition does not change the position of the SPE (a diamond nanocrystal in our case) placed on the ring cavity.

## 2. Theoretical Model

We consider a ridge waveguide with the lower and upper cladding being SiO$_2$ and air, respectively. Figure 1(a) shows the cross-section of the ridge waveguide. The Si$_3$N$_4$ core has a refractive index of 2.087 (measured value at 737 nm) with width $w$, and height $h = 250$ nm. In this work, we consider two widths, $w = 330$ and 480 nm. The geometrical parameters of the waveguides are designed to support single-mode propagation for both TM and TE polarizations of light with low losses.

We assume that the deposition of SiO$_2$ forms a layer with thickness $t$ on top of the lower cladding layer and core layer, parallel to the $x$-axis. This is schematically shown in Figure 1(b). The normal line of the additional layer is parallel to $y$-axis. In reality, there will be a thin layer of SiO$_2$ (with the normal line along the $x$-axis) surrounding the vertical edges of the core. However, a deterministic value for the thickness of such a layer is hard to obtain. Hence, it has been neglected in the theoretical model. The ridge waveguide (before deposition) represents $t = 0$, and another extreme $t = \infty$ represents a channel waveguide (Figure 1(c)). The example of real distribution of SiO$_2$ over Si$_3$N$_4$ waveguide is shown at Figure 1(d) – (e), (see Experimental Procedure).

The waveguide optical properties as a function of $t$ are obtained by solving the two-dimensional time-independent Maxwell's equation [32]. In Figures 2(a), 2(b) and 2(c), we exhibit the optical mode profile of the waveguides ($w = 480$ nm) for the cases of $t = 0$ (ridge), $t = 150$ nm, and $t = \infty$, respectively for both TE and TM polarizations. For TE polarization, the $E_x$ and $H_y$ components are dominant. On the other hand, for TM polarization, $E_y$ and $H_x$ are dominant. The contrast in refractive index between the upper cladding (air) and the core is larger than the contrast between the lower cladding (SiO$_2$) and the core of the ridge waveguide. Therefore, the horizontal symmetry is broken in the optical mode pattern. Moreover, the degree of mode penetration (skin depth) is higher in the lower cladding than in the upper cladding (Figure 2(a)). As $t$ increases, the upper cladding is modified, and the effective contrast in the refractive index is reduced. Consequently, the asymmetry decreases, and the degree of mode penetration in the upper cladding region increases. In the extreme case of $t = \infty$ (channel waveguide), the symmetry is restored.

Figure 3(a) shows the effective refractive index of the optical modes as a function of $t$ for w = 480 nm and λ = 737 nm. The effective refractive index is linked to the group refractive index by the following equation,

$$n_{\text{eff}}(\lambda, t) = n_g(t) + \lambda D(t) \qquad (1)$$

Here both group index $n_g$ and $D = dn_{\text{eff}}/d\lambda$ are also functions of $t$. These parameters can be evaluated as follows. For each $t = t_0$, the effective index for a range of wavelengths in the proximity of λ = 737 nm is calculated. The resulting $n_{\text{eff}}(\lambda, t_0)$ is fitted with a linear line. The regression coefficients are used to estimate $n_g(t_0)$ and $D(t_0)$. This process is repeated for other values of *t* to obtain $n_g(t)$ and $D(t)$. Figures 3(b) and 3(c) graph $n_g(t)$ and $D(t)$, respectively. As we can see from the figure, $n_{\text{eff}}$, $n_g$ and $D$ saturate for large *t*. This is expected because increasing the cladding thickness beyond several skin depths does not affect the confinement of the mode.

If a microring cavity is constructed using a waveguide of effective refractive index $n_{\text{eff}}(\lambda, t)$, then the resonance frequency can be obtained from the resonance condition $m\lambda = n_{\text{eff}}(\lambda, t)L$, where $L = 2\pi R$ is the round–trip length, and *m* is a positive integer. Using this condition, and Eqn. 1, it is straightforward to express $\lambda$ as

$$\lambda(t) = \frac{\frac{L}{m} n_g(t)}{1 - \frac{L}{m} D(t)} \qquad (2)$$

Figures 3(d) and 3(e) graph $\lambda(t)$ for the ring cavity with *R* = 8 µm, for *w* = 330 nm and 480 nm, respectively. The circles in the figure represent the wavelengths calculated by Eqn. 2 with specific values of *m*. We found that this behavior can be succinctly represented by an empirical relationship (solid lines in Figures 3(d) and 3(e)),

$$\lambda(t) = \lambda_\infty - \frac{B}{1 + e^{-at}} \qquad (3)$$

where *B* and *a* are positive constant coefficients, and $\lambda_\infty$ represents the resonance wavelengths for *t* = ∞. The fitted values of *B* and *a* are tabulated in Table 1. Clearly, near t ≈ 0, wavelength varies linearly with *t*. As *t* increases, the rate decreases exponentially and eventually saturates.

## 3    Experimental Procedure

### 3.1    Fabrication

We fabricated ring cavities with a radius (*R)* 8 µm and bus waveguides with identical height (*h*) and width (*w*) at a commercial CMOS foundry. The gap between the bus and the ring is 160 nm. This value is close to the critical gap of the ring resonator that maximizes the bus–ring coupling. For the efficient coupling of light from a lensed optical fiber, each waveguide has 150 nm tapered edge couplers for the efficient transformation of $TEM_{00}$ mode of the lensed fiber into propagating modes of $Si_3N_4$ waveguide.

The fabrication started with a standard 8-inch silicon wafer. A 3.1 µm-thick $SiO_2$ layer and a 250 nm-thick $Si_3N_4$ were formed using the thermal oxidation and low-pressure chemical vapor deposition (LPCVD) method. Before patterning the $Si_3N_4$ layer, chemico-mechanical polishing (CMP) was used to reduce the thickness of the $SiO_2$ to 3 µm. Then 248 nm KrF deep ultraviolet (DUV) lithography and inductively coupled plasma (ICP) etch were used to define the $Si_3N_4$ waveguides and ring resonators. This was followed by the upper cladding deposition of 3.4 µm-thick $SiO_2$ via plasma-enhanced chemical vapor deposition (PECVD). To expose the $Si_3N_4$ ring resonators to the air, a window was opened in the upper cladding layer of $SiO_2$ using lithography and a combination of dry etch (3.1µm) and wet etch (0.3µm). The edge coupler for light coupling with fiber was fabricated by the deep etch of the cladding and Si substrate to realize a trench.

The layers of $SiO_2$ were deposited in steps of 50 nm by ICP-CVD Oxford PlasmaPro System 100 in $SiH_4$ and $N_2O$ atmosphere at 4 mT pressure, 150 C temperature, and 1000 W ICP power. After each deposition, we characterize the shift of resonances. The accuracy of deposition for $SiO_2$ and its distribution around the waveguide and the cavity was studied by making Focused Ion Beam (FIB) cross-cut for cavities with radius 8 microns and waveguide widths 480 and 330 nm, and then studying them under Transmission Electron Microscope (TEM) (Figure 1(d) – (e)).

The nanodiamonds (Sigma-Aldrich) in IPA solvent (0.1 wt%) were ultrasonicated for 30 minutes and then deposited on the chip by spin-coating at 2000 rpm for 5 minutes.

Occasionally, the nanodiamonds were deposited on top of the $Si_3N_4$ cavities. SEM imaging was then performed to locate nanodiamonds on photonic structures.

*3.2 Measurements*

We perform measurements at the wavelengths around 738 nm by probing the chip with either attenuated broadband femtosecond (80 fs) pulses (Mai Tai, SpectraPhysics) or tunable narrowband (less than 1 MHz linewidth) diode laser (Sacher), see Figure 4. Lasers are coupled to the waveguide by a lensed fiber (Oz Optics) and a specially designed coupler at the edge of the chip. The transmitted light was collected with a 50x objective lens (Olympus) at an opposite edge of the chip and then coupled into the single-mode fiber. The signal was then sent to an optical spectral analyzer (OSA, Yokogawa AQ6370) with a spectral resolution of 37 pm in the visible range. With its spectral width of over 10 nm, the fs-laser covered several resonances and allowed fast estimation of the resonance position. The precise measurement of resonances is performed by a narrowband tunable diode laser. We sweep the laser wavelength in the vicinity of the resonance and measure the light intensity passing through the cavity using an optical power meter (Thorlabs) placed after the collimating objective. The wavelength of the tunable laser was controlled by a wavelength meter (HighFinesse, Model WS-7) with an accuracy of up to 60 MHz and resolution of 2 MHz. This allows us to measure the quality factor and fine shifts of the resonance. The experimental data was acquired and post-processed at the PC using self-written software on Python.

**4. Results and Discussions**

*4.1 Demonstration of tuning using $SiO_2$ deposition*

For very thin layers deposited onto the cavity, we cannot precisely characterize the shift, as it is larger than one $FSR = \lambda^2/(2\pi n_{\text{eff}} R)$. The measured Q-factor is at the order of $10^4$ (see Experimental Section), which we confirm to be unaffected by the deposition of the cladding layers, see Supplementary Figure S1. Starting from some finite thickness of $SiO_2$ layer (200 nm in our case), the resonances shift stays within one FSR, and we can track its position with the same order ($m$ in Eq. 2). Figure 5 shows the experimentally measured resonance wavelengths as a function of $t$. The theoretical curves are obtained by averaging the numerical derivatives of the respective empirical fits of the two modes in Figures 3(d) and 3(e). As it can be readily seen from Figure 5, there is a good agreement in the tuning rate between the theoretical predictions and experimental measurements. The discrepancy in the absolute experimental and theoretical values (see Table 1 for the parameters of empirical fits) is due to constraints of the theoretical model. Specifically, it does not include a thin $SiO_2$ layer deposited around the vertical edges of the core (with the normal of the layer parallel to *x*-axis, see Figure 1). The change in the geometry after every deposition occurs predominantly along *y*-axis, rather than *x*-axis. Therefore, a good numerical agreement can be obtained by considering the amount of tuning after every increment in the thickness.

In Figure 5 we can see that every additional layer of $SiO_2$ leads to a decreasing shift in the resonance. In a cavity with $w = 330$ nm and TE polarization, a tuning as large as 3 nm can be obtained when one increases the thickness from 200 nm to 250 nm. In general, the tuning rate is more significant for smaller waveguide widths. Indeed, smaller widths produce optical modes with weaker confinements and thus are more vulnerable to perturbations.

*4.2 Fine-tuning using a laser*

Comparing theory and experimental results, we can see that tuning with $SiO_2$ cladding is not very precise. At the same time, it strongly shifts the cavity resonance. We suggest that our method could be complementary to other existing methods for fine-tuning. Cladding with $SiO_2$ brings the resonance close to the desired wavelength, and then precise tuning can be achieved by other methods. This two-step approach can be useful when tuning requires significant power consumption and/or heat dissipation. One particular scenario is when cavities with nanoparticles containing SPEs are used at cryogenic temperatures. The heat produced by the chip can degrade the properties of SPEs, for example, by causing the drift of the emission line.

We further demonstrate the two-step tuning by depositing 300 nm cladding and heating the cavity by a laser. We use a 532 nm continuous wave laser focused onto the ring cavity by a 100x objective (Nikon, NA=0.9). By adjusting the laser power, we can precisely tune the resonant frequency line of the cavity within the range of 12 pm, see Figure 6. To ensure that the localized heating is responsible for the shift of the cavity resonance, we moved the laser spot by 1 micron away from the cavity. In this case, we observed that the resonance line returns to the initial position. Thus we demonstrate that the cavity can be fine-tuned using the combination of cladding deposition and localized laser heating, without disturbing the SPE itself.

### 4.3 Integration of the nanoparticles

Integration of nanoparticles hosting SPEs with photonic platforms is one of the main targets for the successful realization of quantum photonics. Here we demonstrate that $SiO_2$ deposition does not significantly affect the positioning of nanoparticles deposited on top of photonic structures. We spin-coat nanodiamonds on top of our samples so that with some probability, the nanodiamonds are localized on top of the ring cavity. We then deposit the $SiO_2$ layer of 500 nm, as described above. By comparing Scanning Electron Microscope (SEM) images taken before and after $SiO_2$ deposition, we find that the position of nanodiamonds does not change within the resolution of the SEM, see Figure 7. It shows that our method can be implemented for the case of cavities integrated with nanoparticles.

## 5    Conclusions

We demonstrate a technique for broadband tuning of $Si_3N_4$ microring cavities. The method relies on the deposition of $SiO_2$ nanolayers onto the waveguide cladding and combining it with localized laser heating. Experimental results show good agreement with the theoretical predictions; some discrepancy in the explicit values can be attributed to the non-uniform distribution of $SiO_2$ layers.

Our technique could be complemented by thermal tuning method that provides high accuracy tuning. Thus we are able to tune the cavity resonance without significant heat dissipation, which is important for operation at cryogenic temperatures. In contrast to the application of photochromic or polyelectrolyte films, we do not observe the degradation of the Q-factor. Furthermore, we verify that nanoparticles placed on the cavity are not displaced by the deposition of $SiO_2$.

Our technique can be helpful for small-volume, high Purcell-factor cavities with large FSRs, when the resonance and target wavelengths can be separated by several nanometers. The demonstrated technique will be a valuable addition to the toolkit for developing integrated quantum photonic devices.

**Acknowledgements**

We acknowledge the support of the Quantum Technology for Engineering (QTE) program of A*STAR project A1685b0005 and project C210917001.

**Author contributions:** D.A.K, G.A. and L.A.K. jointly conceived the idea of the experiment. G.A. and C.E.P. designed the optical devices. T.H. and N.L. fabricated the samples. D.A.K. and V.L. build the chip characterization setup and conducted optical measurements. G.A. and D.A.K. analysed the experimental data. D.A.K. wrote the first draft of the manuscript with the contributions of all co-authors. D.A.K., G.A. and L.A.K. coordinated the project.

**Conflict of interest statement:** The authors declare no conflicts of interest regarding this article.

**Table 1: The values of *B* and *a* in the empirical relationship of Eqn. 3. Mode 1 has higher *m* than mode 2.**

| *w*=330 nm | TE | | TM | |
|---|---|---|---|---|
| | *B* | *a* | *B* | *a* |
| Expt. | 0.20 | 18.03 | 0.13 | 17.72 |
| Theory (mode 1) | 0.10 | 14.08 | 0.05 | 10.60 |
| Theory (mode 2) | 0.10 | 14.15 | 0.05 | 10.64 |
| *w*=480 nm | TE | | TM | |
| | *B* | *a* | *B* | *a* |
| Expt. | 0.04 | 16.22 | 0.05 | 18.03 |
| Theory (mode 1) | 0.04 | 14.49 | 0.03 | 11.41 |
| Theory (mode 2) | 0.04 | 14.55 | 0.03 | 11.45 |

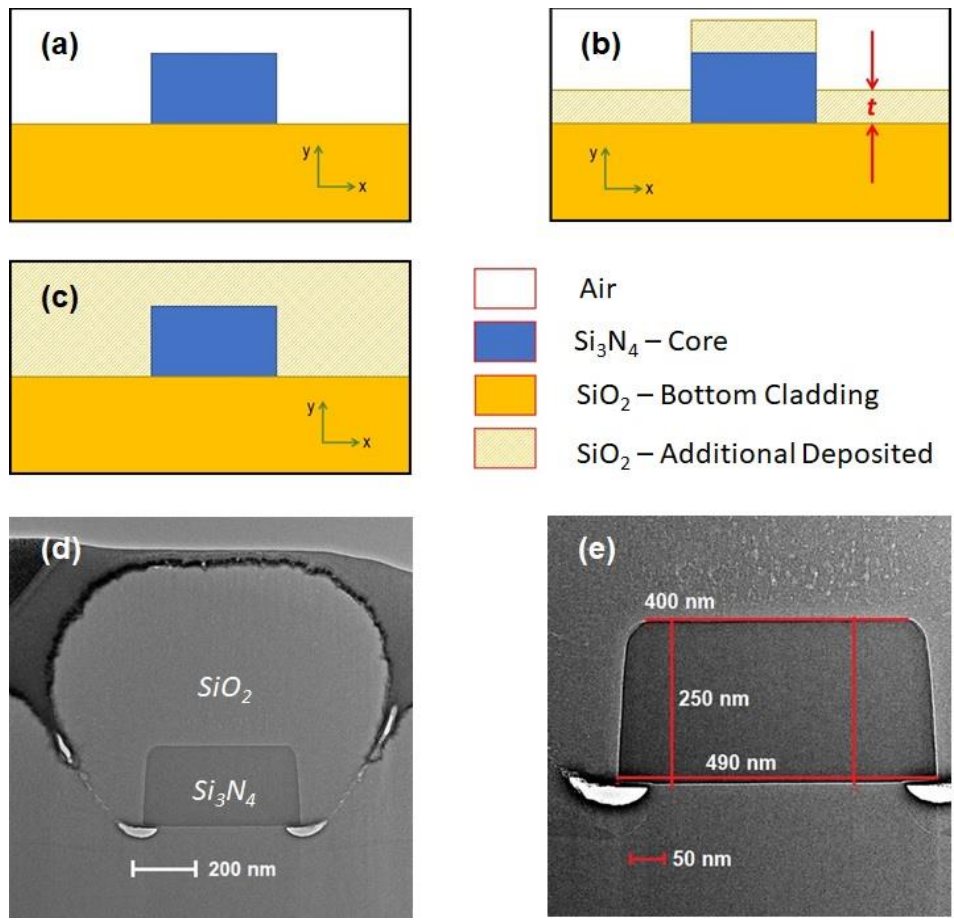

**Fig. 1:** Distribution of SiO$_2$ cladding around Si$_3$N$_4$ waveguide **(a)** Schematic of a ridge waveguide with lower and upper cladding being air and SiO$_2$, respectively. **(b)** Schematic for theoretical consideration with an additional layer of SiO$_2$ deposited. **(c)** A channel waveguide, with ideally the upper and lower cladding being SiO$_2$. **(d)** FIB crosscut for 480 nm Si$_3$N$_4$ waveguide and SiO$_2$ cladding measured by TEM. **(e)** Geometrical dimensions of 480 nm Si$_3$N$_4$ waveguide

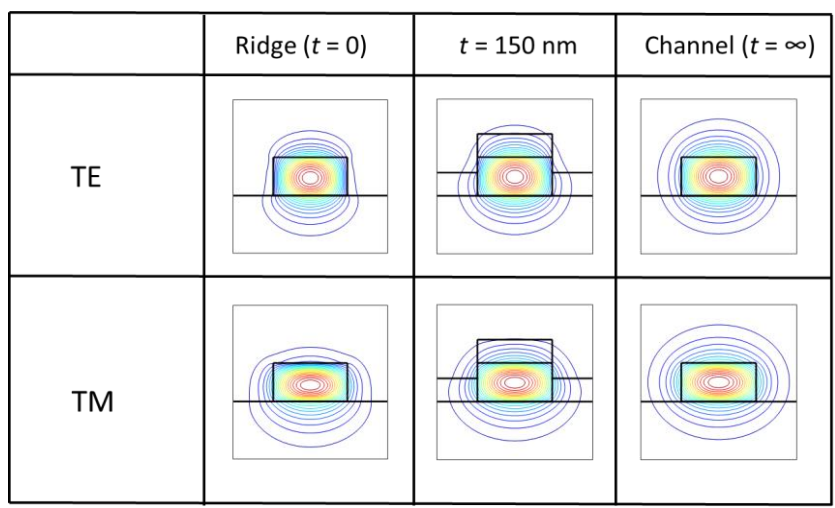

**Fig. 2:** Simulation results for the contour plots of the optical mode ($H$ – field) for the waveguide with $h = 250$ nm, $w = 480$ nm. The red (blue) color represents the maxima (minima).

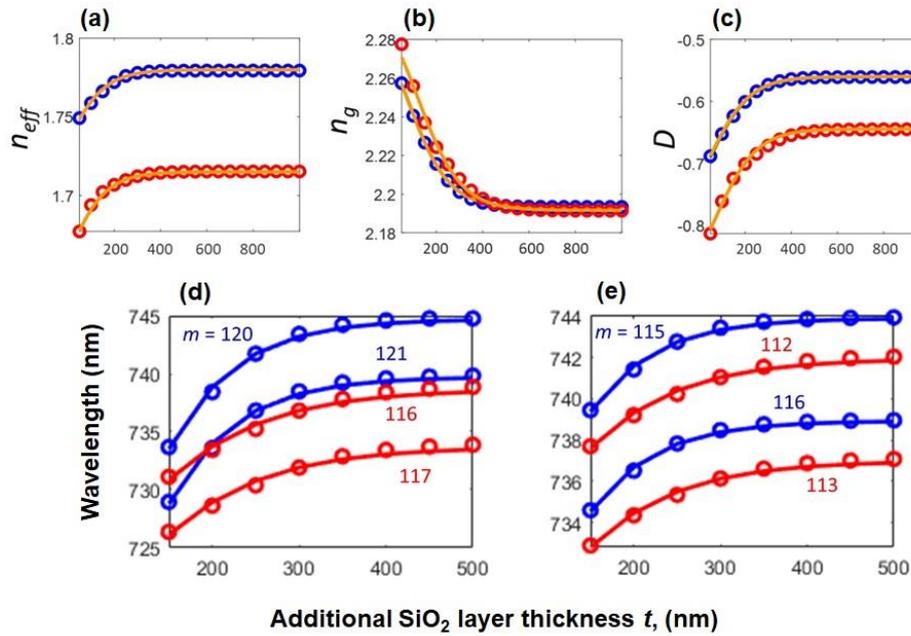

**Fig. 3:** The theoretically estimated optical properties as a function of additional SiO$_2$ layer thickness, red and blue colors represent TE and TM polarizations, respectively: **(a)** Effective refractive index, **(b)** Group Refractive Index, and **(c)** the coefficient $D$ for the waveguide with $h = 250$ nm, $w = 480$. **(d)** and **(e)** Theoretically estimated resonance wavelengths of the ring cavity ($R = 8$ µm) for the cavity widths **(d)** 330 nm, and **(e)** 480 nm, respectively. Circles are obtained from Eqn. 2, while the solid lines are empirical fits given by Eqn. 3.

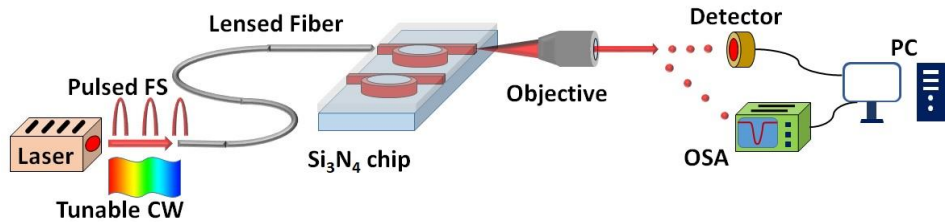

**Fig. 4:** Sketch of the experimental setup: light from broadband pulsed fs laser or tunable CW laser is coupled to the lensed fiber. Lensed fiber injects the light into the Si$_3$N$_4$ chip with ring cavities. At the output of the chip the transmitted light is collected and collimated with 50x objective. Then the light is sent either to the photodetector (in case of tunable CW laser) or to the optical spectral analyzer (OSA) (in case of FS laser).

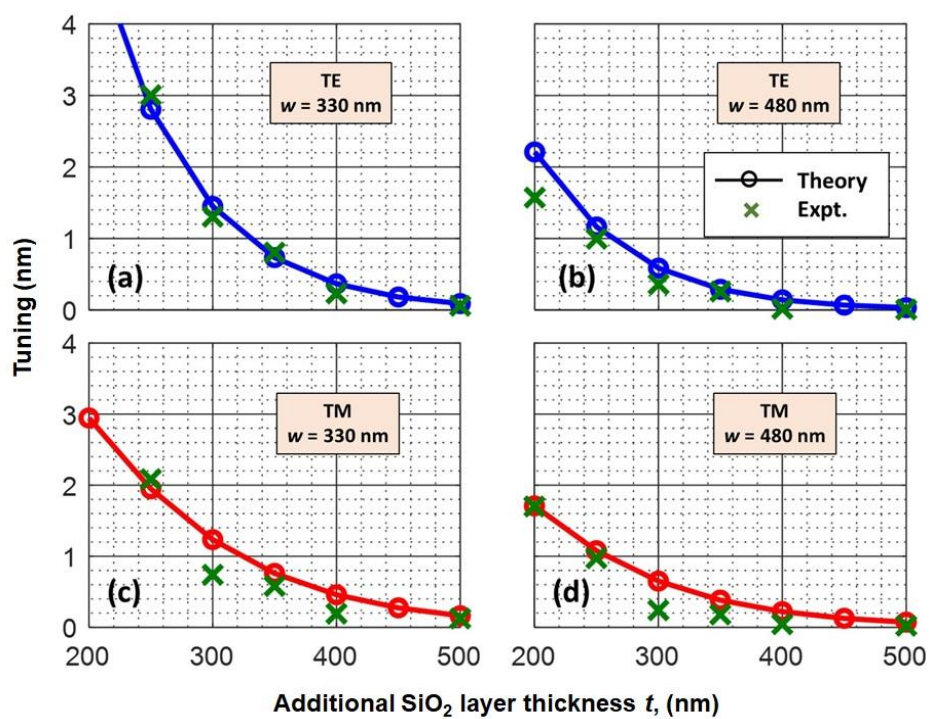

**Fig. 5:** Experimental results for tuning the resonance wavelength as a function of additional SiO$_2$ layer thickness. The experimental points are plotted as crosses, while solid lines represent theoretical curves. Red and blue colors represent TE and TM polarizations, respectively.

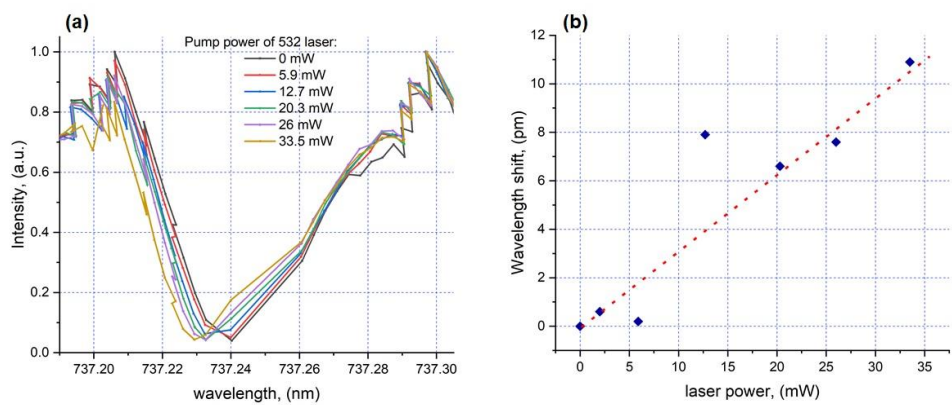

**Fig. 6:** Experimental tuning of the resonance wavelength for Si$_3$N$_4$ ring cavity clad with 300 nm of SiO$_2$ by local heating with 532 nm laser. (**a**) Spectral profiles of the resonances for different pump powers of heating laser; (**b**) the dependence of reversible shift of the central resonance wavelength on applied pump powers of heating 532 nm laser. The red dotted line is a fit indicating the linear dependence of the refractive index (tuning) on temperature.

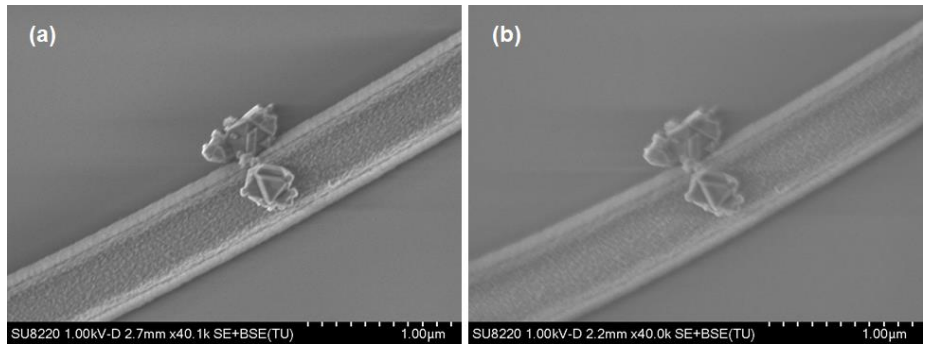

**Fig. 7:** SEM picture of nanodiamonds placed on top of $Si_3N_4$ ring cavity before **(a)** and after **(b)** $SiO_2$ deposition of 500 nm thickness.

*Supplementary Materials*

Here we compare resonances before and after deposition of an additional layer of $SiO_2$. As it can be seen from the figures below cladding with $SiO_2$ does not cause degradation of the Q-factors of resonances in contrast with the application of photochromic or polyelectrolyte films.

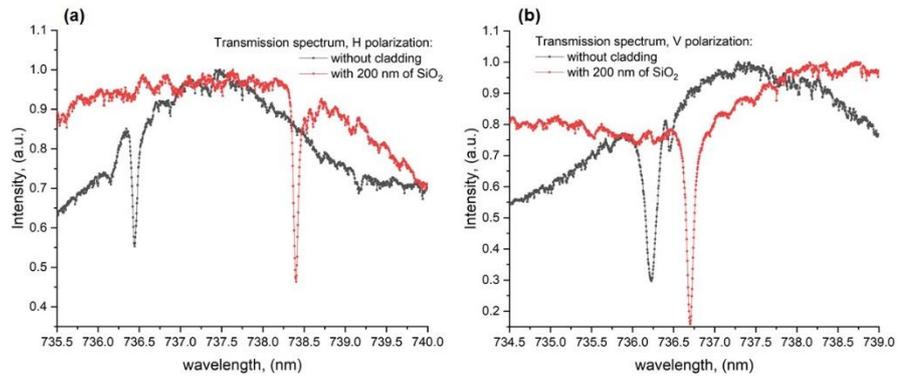

**Fig. S1:** Transmission spectrum of $Si_3N_4$ ring cavity with R=8 µm and $w$=480 nm measured with OSA for horizontal (TM) **(a)** and vertical (TE) **(b)** polarizations of the input light. Black denotes results before cladding with $SiO_2$, red stands for results after deposition of 200 nm of $SiO_2$.